\documentstyle{article}
\newtheorem{theorem}{Theorem}[section]

\newenvironment{prooof}{\begin{description}
                   \item[{\small {\bf Proof:}}] \small}{\hfill {\bf Q.E.D.}
                                                          \medskip
                                                       \end{description}}

\newtheorem{defi}{Definition}[section]
\newtheorem{prop}{Proposition}[section]
\newtheorem{lemma}{Lemma}[section]
\newtheorem{rem}{Remark}[section]

\newcommand{\bdef}{\begin{defi}}
\newcommand{\ede}{\end{defi}}
\newcommand{\bsat}{\begin{theorem}}
\newcommand{\esat}{\end{theorem}}
\newcommand{\bprop}{\begin{prop}}
\newcommand{\eprop}{\end{prop}}
\newcommand{\blem}{\begin{lemma}}
\newcommand{\elem}{\end{lemma}}
\newcommand{\brem}{\begin{rem}}
\newcommand{\erem}{\end{rem}}
\newcommand{\bbew}{\begin{prooof}}
\newcommand{\ebew}{\end{prooof}}
\newcommand{\be}{\begin{equation}}
\newcommand{\ee}{\end{equation}}
\newcommand{\bea}{\begin{eqnarray}}
\newcommand{\eea}{\end{eqnarray}}
\newcommand{\beas}{\begin{eqnarray*}}
\newcommand{\eeas}{\end{eqnarray*}}
\newcommand{\ben}{\begin{enumerate}}
\newcommand{\een}{\end{enumerate}}
\newcommand{\lb}{\label}

\newcommand{\ra}{\rightarrow}
\renewcommand{\L}{{\cal L}}

\newcommand{\f}{\frac}
\newcommand{\p}{\partial}

\newcommand{\Real}{\mbox{I \hspace{-0.82em} R}}

\newcommand{\Nf}{{\cal N}}

\newcommand{\vp}{\varphi}

\begin{document}
 \begin{titlepage}

  \begin{center}
    {\Large Diffeomorphism Invariant Integrable Field Theories} \\
       \vspace{0.2cm}
                       {\Large and} \\
       \vspace{0.2cm}
      {\Large Hypersurface Motions in Riemannian Manifolds} \\
             \vspace{1cm}
     {\bf Martin Bordemann} \\
         Fakult\"at f\"ur Physik \\
        Universit\"at Freiburg \\
          Hermann-Herder-Str. 3\\79104 Freiburg i.~Br., F.~R.~G \\
          e-mail: mbor@phyq1.physik.uni-freiburg.de \\
        \vspace{0.3cm}
     {\bf Jens Hoppe} \footnote{Heisenberg Fellow\\~~~On leave of absence
                                 from Karlsruhe University}\\
      Institut f\"ur Theoretische Physik\\
       ETH H\"onggerberg \\
       CH 8093 Z\"urich, Switzerland \\
           e-mail: hoppe@itp.phys.ethz.ch \\
      \vspace{0.3cm}
               FR-THEP-95-26 \\
            ETH-TH/95-31   \\
             November 1995\\
             hep-th/9512001
   \end{center}
      \vspace{0.5cm}
     {\small
      \begin{center}
          {\bf Abstract}
      \end{center}
        We discuss hypersurface motions in Riemannian manifolds whose normal
velocity is a function of the induced hypersurface volume element
and derive a second order partial differential equation for the
corresponding time function $\tau(x)$ at which the hypersurface passes
the point $x$. Equivalently, these motions may be described in a Hamiltonian
formulation as the singlet sector of certain diffeomorphism invariant
field theories. At least in some (infinite class of) cases,
which could be viewed as a large-volume limit of Euclidean $M$-branes
moving in an arbitrary $M+1$-dimensional Riemannian manifold,
the models are integrable:
In the time-function formulation the equation becomes linear
(with $\tau(x)$ a harmonic function on the embedding
Riemannian manifold). We explicitly compute solutions
to the large volume limit of Euclidean membrane dynamics in $\Real^3$
by methods used in electrostatics and point an additional gradient
flow structure in $\Real^n$.
In the Hamiltonian formulation we discover
infinitely many hierarchies of integrable, multidimensional, $N$-component
theories possessing infinitely many diffeomorphism invariant, Poisson
commuting, conserved charges.}
\end{titlepage}

{\bf 1.} In physics, and mathematics, surface motions are usually
considered independent of the parametrisation, the most prominent
example perhaps being the `flow by mean curvature' which has been
of intense mathematical interest \cite{stern} as well as
of physical importance \cite{sternstern}. When studying the dynamics of
relativistic membranes, on the other hand, one encounters \cite{BH94}
in a partially gauge-fixed formulation, a very interesting equation
(describing the time-evolution of a 2-dimensional surface in $\Real^3$)
which is parametrisation-{\em dependent} (meaning that different
parametrisations of the initial surface $\Sigma_0$ generically lead to
geometrically different shapes $\Sigma_t$ at later times, $t$
(in the case at hand there is a residual symmetry group of area-preserving
diffeomorphisms, but any reparametrisation whose Jacobi determinant is
different from $1$ will generically change the motion);
alternatively, the equation (though first order in time) may be viewed
as requiring the initial unparametrized shape $\Sigma_0$ {\em and} the
initial unparametrized normal velocity field on $\Sigma_0$ as initial
conditions (the parametrisation of $\Sigma_0$, and of $\Sigma_t$,
then follows, up to area-preserving diffeomorphisms of $\Sigma_0$
-that don't change the geometry of the motion- from the equation).
Moreover, it was
shown in \cite{YITP} that this seemingly parametrisation dependent
equation can be obtained from the singlet sector of a diffeomorphism
invariant Hamiltonian field theory. In \cite{Hop94}, on the other hand,
some surface motions in $\Real^3$ were shown to be best described
by a (second order partial differential) equation for $\tau (\vec{x})$,
the time at which $\Sigma_t$ contains $\vec{x}$.

In this letter, we significantly generalize this situation (to
higher dimensions, curved embedding spaces, and more general
dynamics) by considering a rather general class of parametrisation
dependent hypersurface motions in Riemannian manifolds. For these
hypersurface motions
we shall give an equivalent diffeomorphism-invariant
(constrained) Hamiltonian formulation
as well as derive a second order partial differential equation for
the time-function $\tau(x)$, which in some cases (that may be viewed
as large-volume limits of Euclidean $M$-branes or, equivalently,
as highly generalized, multilinear, field theoretic extensions of
Nahm's matrix equations \cite{what?}) turns out to simply be
$\Delta \tau =0$
where $\Delta$ is the Laplacean of the embedding Riemannian manifold
in which the motion takes place.
In the particular case of three-dimensional Euclidean space the
solution of the Laplace equation for the time function becomes
easy to handle thanks to the formal equivalence with the potential equation
of electrostatics: we compute explicit solutions of these surface motions
determined by singular one-dimensional membranes at $t=-\infty$ and
$t=+\infty$
(corresponding to negatively and positively charged composed loops of wire,
respectively) between which two-dimensional surfaces develop in time.

Finally, and perhaps most importantly, we are led to an infinite set of
hierarchies of multidimensional
integrable diffeomorphism invariant $N$-component field theories
for which we are able to provide a complete set of Poisson-commuting,
diffeomorphism invariant, conserved charges.

\vspace{0.4cm}

{\bf 2.} Let $\Sigma$ be an orientable compact connected manifold of
dimension
$M$ with a fixed volume form $\rho$. We shall denote co-ordinates on
$\Sigma$ by $\varphi^r,1\leq r\leq M$. Let $({\cal N},\zeta)$ be an
orientable
Riemannian manifold of dimension $N=M+1$ on which we shall denote
co-ordinates by $x^i,1\leq i\leq M+1$ and whose volume form induced by
$\zeta$ will be denoted by $\omega_\zeta$. Consider the set of smooth
immersions
$x:\Sigma\ra{\cal N}$, $imm(\Sigma,{\cal N})$. For each immersion $x$
there is a unique outward normal field $n[x]:\Sigma\ra T{\cal N}$
(depending on $x$ and its first derivatives) attached
to the parametrized hypersurface $x:\Sigma\ra{\cal N}$. Moreover, let
$x^*\zeta$ denote the Riemannian metric on $\Sigma$ obtained by
pulling back $\zeta$ to $\Sigma$ by $x$ and denote by $\omega_{x^*\zeta}$
the corresponding volume form on $\Sigma$. It follows that there is a
unique smooth positive function $\sqrt{g[x]}/\rho:\Sigma\ra\Real$ depending
on $x$, its first derivatives, and the reference volume form $\rho$
such that $\omega_{x^*\zeta}=(\sqrt{g[x]}/\rho)\rho$. Finally, let
$\alpha$ be a positive smooth fuction on (a suitable open interval of)
the positive real line with nowhere vanishing derivative. Consider
now the hypersurface motion $(-\epsilon,\epsilon)\ra imm(\Sigma,{\cal N}):
t\mapsto x(t) $ in ${\cal N}$ described by the following
differential equation:
\be \label{surf}
  \dot{x}~:=~\f{\p x}{\p t}~=~\alpha(\sqrt{g[x]}/\rho)~n[x]\quad .
\ee
In local co-ordinates this equation looks as follows: writing
\be g_{rs}=\f{\p x^i}{\p \varphi^r}\f{\p x^j}{\p \varphi^s}\zeta_{ij}(x)
                                       \label{smallg} \ee
for the induced metric $x^*\zeta$ we get
\begin{eqnarray}
  \dot{x}^i & = & \alpha(\sqrt{\det(g_{rs})}/\rho) \nonumber \\
            &   &
     \f{1}{M!\sqrt{\det(g_{rs})}}\zeta^{ij}(x)\sqrt{\det(\zeta_{kl})}(x)
           \epsilon_{ji_1\cdots i_M}\epsilon^{r_1\cdots r_M}
            \f{\p x^{i_1}}{\p \varphi^{r_1}}\cdots
              \f{\p x^{i_M}}{\p \varphi^{r_M}}  \label{surfexplicit}
\end{eqnarray}
The following choices of $\alpha$ are of particular interest:
\begin{eqnarray}
  {\rm Lorentzian~}M{\rm -brane} & : &
              \dot{x}=\sqrt{1-(\sqrt{g[x]}/\rho)^2}~n[x] \label{Lorentz} \\
  {\rm Euclidean~}M{\rm -brane}  & : &
              \dot{x}=\sqrt{(\sqrt{g[x]}/\rho)^2-1}~n[x] \label{Euklid}
\eea
and the large-volume limit of (\ref{Euklid}),
\be
        \dot{x}=(\sqrt{g[x]}/\rho)~n[x]. \label{Nahm}
\ee
The small-volume limit of (\ref{Lorentz}),
\be \lb{Optical}
            \dot{x}=n[x]
\ee
(which may be called the {\em Optical model})
is not contained in this class (since the prefactor $\alpha$ of the normal
is constant). Nevertheless, it can be solved directly since (\ref{Optical})
is easily seen to imply the following equation of free motion:
\be
  \f{\p^2x^i}{\p t^2}+\Gamma^i_{jk}(x)\f{\p x^j}{\p t}\f{\p x^k}{\p t}=0
\ee
(where $\Gamma^i_{jk}$ denote the Christoffel symbols of $\zeta$) which can
be solved as soon as the geodesic flow of $(\Nf,\zeta)$ is
explicitly known (e.g. for flat $\Real^N$, the $N$-sphere, etc.).
Some of the following results will also remain true for (\ref{Optical}).

As in \cite{Hop94} we are trying to derive a differential equation for
the time-function:

\bsat \label{timefunction}
Let $\Sigma$, $\rho$, $\Nf$, $\zeta$ and $\alpha$ be defined as above.
\begin{enumerate}
 \item Suppose that there is a positive real
   number $\epsilon$ and a solution
   $x$ of the above equation (\ref{surf}) such that the map
   $x:(-\epsilon,\epsilon)\times\Sigma\ra{\cal N}:
   (t,\varphi)\mapsto x(t,\varphi)$ is a diffeomorphism onto its image,
   ${\cal N}_\epsilon$,
   which is an open neighbourhood of the initial hypersurface $x(0,\Sigma)$.

   Then the first component of the inverse map
   $x^{-1}:{\cal N}_\epsilon\ra (-\epsilon,\epsilon)\times\Sigma$
   which we shall call the {\em time function} $\tau$ satisfies
   the following equations where $\nabla\tau:=\zeta^{\#}(d\tau)$
   ($\nabla^i\tau=\zeta^{ij}\p_j\tau$) denotes the gradient of $\tau$,
   $|v|:=\sqrt{\zeta(v,v)}=\sqrt{\zeta_{ij}v^iv^j}$ for any tangent vector
   $v$ to $\cal N$,
   and the symbol ``;'' stands for covariant derivative.
   \begin{eqnarray}
             n[x]  & = & \f{\nabla \tau}{|\nabla\tau|}(x)
                                               \label{normgrad} \\
   |\nabla\tau|(x) & = & \f{1}{\alpha(\sqrt{g[x]}/\rho)},
                                         \label{scalarconstraint} \\
                 0 & = & |\nabla\tau|^2\zeta^{ij}\tau_{;ij}
                       ~+~(\tilde{\beta}(|\nabla\tau|)-1)
                                   \tau_{;ij}\nabla\tau^i\nabla\tau^j.
                                     \label{taueqn}
   \end{eqnarray}
   where
   $\tilde{\beta}$ is defined by:
   \be
      \tilde{\beta}(z)~:=~-z~\f{\p}{\p z}(\alpha^{-1}(1/z)).
   \ee
   Eqs (\ref{normgrad}) and (\ref{scalarconstraint}) remain true for
   positive $\alpha$ whose derivative may vanish.

  \item Conversely, suppose that the smooth function
   $\tau:\Nf\ra\Real$ is a solution of the second order partial differential
   equation (\ref{taueqn}) and that there is a positive real number
   $\epsilon$ such that all its level surfaces $\tau =c$ for
   $c\in (-\epsilon,\epsilon)$ are diffeomorphic to $\Sigma$.

   Then there is a parametrisation
   of the zero level surface $\Sigma_0$ of $\tau$, $x_0:\Sigma\ra\Sigma_0$
   and a positive real number $r$ such that
   eqn (\ref{scalarconstraint}) is satisfied for $x_0$ and $r\rho$.
   Moreover,
   let $\Phi_t$ denote the flow of the vector field
   $X:=\nabla\tau/|\nabla\tau|^2$, i.e. the solution of the ordinary
   differential equation $\p\Phi_t(x)/\p t=X(\Phi_t(x))$ with initial
   condition $\Phi_0(x)=x$. Then for each parametrisation $x_0$ of the
   zero level surface which satisfies the above condition
   (\ref{scalarconstraint}) the map $x:(-\epsilon,\epsilon)\times\Sigma
   \ra\Nf$ defined by
   \be
     x(t,\vp):=\Phi_t(x_0(\vp))
   \ee
   satisfies equation (\ref{surf}) (with $\rho$ replaced by $r\rho$) with
   initial condition $x_0$.
 \end{enumerate}
\esat
\bbew
1. It is obvious that the gradient of $\tau$
will be orthogonal on the
surfaces of constant time whence the gradient of $\tau$ is proportional
to the surface normal $n[x]$ which gives (\ref{normgrad}).
Using the inverse function theorem
\[ \delta^i_j=\f{\p x^i}{\p t}\f{\p \tau}{\p x^j}+
               \f{\p x^i}{\p \varphi^r}\f{\p \varphi^r}{\p x^j}, \]
contracting with $\zeta_{ik}n[x]^k$, and using equation (\ref{surf})
and the definition of the surface normal we get the second equation
(\ref{scalarconstraint}). This leads to
\be
   \dot{x} =  \f{\nabla \tau}{|\nabla\tau|^2}(x) \label{newequat}
\ee
In order to obtain a differential equation for the time function
$\tau$ in the case where the derivative of the function $\alpha$ does
never vanish we differentiate equation
(\ref{scalarconstraint}) with respect to
time and use equation (\ref{newequat}) to get rid of $\dot{x}$. Thereby
the time derivative of the left hand side of (\ref{scalarconstraint})
can solely be expressed by second covariant derivatives of the time function.
To do the same for the right hand side we use the equation
\[ \dot{\sqrt{g[x]}}=\f{1}{2}\sqrt{g[x]}g^{rs}\dot{g}_{rs}, \]
obtain by the chain rule
\[ \p_r\dot{x}^i = \p_j(\nabla\tau^i/|\nabla\tau|^2)\p_rx^j  \]
observe that the orthogonal projection $\pi$ onto the tangent spaces
of the hypersurfaces can be expressed in two ways by
\[ g^{rs}\p_rx^i\p_sx^j=\pi^{ij}(x)=\zeta^{ij}(x)-n[x]^i n[x]^j  \]
(see e.g. \cite{BH94}), and replace the normal vectors by the
normalized gradients
of the time-function (\ref{normgrad}) which gives the second order
equation (\ref{taueqn}).

2. Denote by $\Sigma_0$ the level surface $\tau=0$ and by
$i:\Sigma_0\ra\Nf$ the canonical inclusion. Let $\omega_{i^*\zeta}$
denote the volume form on $\Sigma_0$ induced by the metric $\zeta$.
Consider the modified volume form
\[
  \f{1}{\alpha^{-1}(\f{1}{|\nabla\tau|(i)})}\omega_{i^*\zeta}
\]
on $\Sigma_0$. Let $r$ be the unique positive real number such that
the two following integrals are equal:
\[
  \int_{\Sigma}\!r\rho~=~\int_{\Sigma_0}\!
                     \f{1}{\alpha^{-1}(\f{1}{|\nabla\tau|(i)})}
                                  \omega_{i^*\zeta}
\]
By Moser's lemma (see \cite{Mos65}) there is a diffeomorphism
$x_0:\Sigma\ra\Sigma_0$ such that the two volume forms are diffeomorphic,
i.e.:
\[
   r\rho~=~x_0^*(\f{1}{\alpha^{-1}(\f{1}{|\nabla\tau|(i)})}
                                  \omega_{i^*\zeta})
        ~=~  \f{1}{\alpha^{-1}(\f{1}{|\nabla\tau|(x_0)})}
                                  \omega_{x_0^*\zeta}
\]
whence eqn (\ref{scalarconstraint}) is satisfied for $t=0$ and $r\rho$.

Observe now that for all points $y\in\Sigma_0$ we have
\[
  \tau(\Phi_t(y))=t
\]
which can be seen by differentiating the left hand side with respect
to $t$ and showing it to be equal to $1$. When we insert $y=x_0(\vp)$ in this
equation and differentiate with respect to $\vp^r$ we see that
the gradient of the time function is always orthogonal to the surfaces
$\Phi_t(x_0(\Sigma))$. Hence $\nabla\tau(x(t,\vp))/|\nabla\tau(x(t,\vp))|$
is equal to the hypersurface normal $n[x_t](\vp))$ whence we have
the differential equation
\[
  \dot{x}(t,\vp)=\f{\nabla\tau(x(t,\vp))}{|\nabla\tau(x(t,\vp))|^2}
                =\f{1}{|\nabla\tau(x(t,\vp))|}n[x_t](\vp)
\]
We have already shown above that the function (for fixed $\vp\in\Sigma$)
\[
  f(t):=\alpha(\sqrt{g_{rs}(t,\vp)}/(r\rho(\vp)))~|\nabla(x(t,\vp))|
\]
equals $1$ for $t=0$ by construction of $x_0$. Upon differentiating
$f$ with respect to $t$, eliminating $\dot{x}$ by the equation above and
using the second order equation (\ref{taueqn}) we get the first order
time dependent differential equation for $f$:
\[
  \dot{f}(t)~=~
\f{\tau_{;ij}\nabla\tau^i\nabla\tau^j\tilde{\beta}(|\nabla\tau|)}{|\nabla\tau|^4}(x(t,\vp))
               \big( \f{1}{\tilde{\beta}(|\nabla\tau|)(x(t,\vp))}
   -\f{1}{\tilde{\beta}(\f{|\nabla\tau|(x(t,\vp))}{f(t)})}\big)f(t)
\]
Since the right hand side smoothly depends on $f$ in a neighbourhood of
$f=1$ the
obvious constant solution $f(t)=1$ is unique. This proves eqn
(\ref{scalarconstraint}) for all $t$.

\ebew

In other words, the open neighbourhood ${\cal N}_\epsilon$ will be
foliated by the level sets of $\tau$ which are nothing else but the
surfaces of constant time.
Eqs (\ref{normgrad}) and (\ref{newequat}) have also been considered in
\cite{NHW95} and \cite{NW95} and only use
the fact that $\dot{x}$ is proportional to the unit normal.

Note that for nonconstant $\alpha$ the above
surface motion is {\em not} reparametrisation invariant, which is here
reflected in the fact that the time function satisfies a second order
partial differential equation whose normal derivative at the zero
level surface encodes the ``volume part'' of the parametrisation, i.e.
the induced volume form compared to a reference volume. On the
other hand, for the Optical model (\ref{Optical})
the function $\alpha$ is equal to one:
this model of surface motion is invariant under reparametrisation,
and the differential equation
for the time-function is first order, viz. eqn (\ref{scalarconstraint})
for $\alpha=1$
which is the eikonal equation $|\nabla\tau|^2=1$ known in optics.

The second order equations for all the time-functions dealt with in
Theorem \ref{timefunction} can be derived from a Lagrangean:
define the real-valued smooth function $F$ as any solution of the first
order differential equation:
\be \label{Feqn}
     F'(z)=\f{C}{\alpha^{-1}(\f{1}{z})}
\ee
where $C$ is an arbitrary nonzero real number.
Define the Lagrangean $\L$ for the time-function $\tau$ as:
\be
   \L(\p\tau)~:=~\sqrt{\det(\zeta_{ij})}~F(|\nabla\tau|)
\ee
A straight forward calculation shows that the Euler Lagrange equations
produce (\ref{taueqn}) for all $F$ solving (\ref{Feqn}).

For the model (\ref{Nahm}) we can choose the Lagrangean
$\L(\p\tau)$ equal to
$\sqrt{\det(\zeta_{ij}}\f{1}{2}|\nabla\tau|^2$ and get the Laplace equation
\be \label{harm}
     \Delta\tau~=~0
\ee
where $\Delta$ is the Laplacean operator
$\f{1}{\sqrt{\det(\zeta_{ij})}}\p_i(\sqrt{\det(\zeta_{ij})}\zeta^{ij}\p_j$.
This means that -despite its complicated nonlinear appeal- the
model (\ref{Nahm}) is integrable at least for all those
Riemannian manifolds for which the Laplace equation can be solved in a
sufficiently explicit way, e.g. for $\Real^n$, the sphere $S^n$ and
hyperbolic space $H^n$.
We shall discuss the surface motion in $\Real^3$ (for which a linearizability
was already noted in \cite{War90} without using the time function)
according to its harmonic time-function in more detail in the next section.

In numerical relativity theory the foliation of a given spacetime into
spacelike hypersurface which are level surfaces of a (Lorentz) harmonic
time-function has been dealt with by C.~Bona and J.~Mass\'o in
\cite{BM88}. This method of `harmonic slicing' avoids the occurrence
of spacetime singularities on these surfaces.

For the Lorentzian membrane we can choose the Lagrangean equal to the function
$\sqrt{\det(\zeta_{ij})}\sqrt{|\nabla\tau|^2-1}$ and get
the Euler-Lagrange equations
\be
   (1-|\nabla\tau|^2)\Delta\tau+\tau_{;ij}\nabla\tau^i\nabla\tau^j~=~0
\ee
whereas for the Euclidean membrane we can choose
$\L(\p\tau)$ equal to the function
$\sqrt{\det(\zeta_{ij})}\sqrt{1+|\nabla\tau|^2}$
and get the equation
\be
   (1+|\nabla\tau|^2)\Delta\tau-\tau_{;ij}\nabla\tau^i\nabla\tau^j~=~0.
\ee

Note that the global description of surface motion by parametrisation may
drastically differ from the description by the level sets of a time function:
although being equivalent for short time intervals
(and some technical assumptions, see Thm \ref{timefunction}) the former
allows the membrane to be at the same position at different times, but
fixes the topological type, whereas the latter allows for varying topological
type, but every point in space is contained in at most one level surface.
In both pictures one has to admit violations of the regularity of the
mappings (i.e. points where the Jacobians cease to have maximal rank)
in order to produce interesting situations.

\vspace{0.4cm}

{\bf 3.} In this section we should like to discuss some solutions of the
model (\ref{Nahm}) in $\Real^3$ in terms of its harmonic time-function.
A comparison with the Euclidean membrane model (\ref{Euklid})
shows that (\ref{Nahm}) is an approximation of the
Euclidean membrane model in the regime where the induced spatial membrane
area element $\sqrt{\det(g_{rs})}$ is large compared to a reference area
element $\rho$.

The Laplace equation (\ref{harm}) for flat $\Real^3$ can be treated by
methods known in electrostatics
(see e.g. \cite{Jack75}). For example, consider a piece of wire
of length $L$ which is uniformally charged with line charge
$\lambda=q/L$ and which is lying along the $z$-axis between $-p:=L/2$ and $p$:
its electrostatic potential $\tau$ is given by
\be \label{wireeqn}
   \tau(\vec{x})=\lambda\log\big(
                \f{\sqrt{x^2+y^2+(z+p)^2}+z+p}{\sqrt{x^2+y^2+(z-p)^2}+z-p}
                              \big)
\ee
This function is harmonic outside the interval $[-p,p]$ along the
$z$-axis on which it becomes $\pm\infty$ according to the sign
of $\lambda$
and tends to zero for large distances from the origin.
It is known that all the equipotential surfaces of this function
are axially symmetric ellipsoids whose focal points are the end-points
of the piece of wire. In case $\lambda$ is negative we get the -at least
in electrostatics- unusual picture of the equipotential surfaces as
surfaces of constant time: for $\tau=-\infty$ it is a singular line
which blows up into bigger and bigger ellipsoids, and at $\tau=0$ the level
surface is the sphere at spatial infinity.

This situation can be generalized a bit: let
\be
  \vec{x}^-_i,\vec{x}^+_j:S^1\ra \Real^3,~~~~1\leq i\leq m,~~1\leq j\leq n
\ee
be closed curves in $\Real^3$ which are either points or immersed
with only a finite number of
self-intersection points. Suppose furthermore that any two different curves
do not intersect. Moreover, let $\lambda^-_i,\lambda^+_j:S^1\ra\Real$
be positive smooth functions where $1\leq i\leq m,~~1\leq j\leq n$.
On $\Real^3$ minus all the images of the closed curves,
$\vec{x}^-_1(S^1),\ldots,\vec{x}^-_m(S^1),\vec{x}^+_1(S^1),\ldots,
\vec{x}^+_n(S^1)$ we can define the following harmonic function $\tau$:
\be \label{Nahmsolution}
  \tau(\vec{x}):=-\sum_{i=1}^m\int_0^{2\pi}\!d\vp
                 \f{\lambda^-_i(\vp)}{|\vec{x}-\vec{x}^-_i(\vp)|}
                 +\sum_{j=1}^n\int_0^{2\pi}\!d\vp
                 \f{\lambda^+_j(\vp)}{|\vec{x}-\vec{x}^+_j(\vp)|}
\ee
(It is harmonic because $\vec{x}\mapsto\f{1}{|\vec{x}-\vec{y}|}$ is harmonic
for all $\vec{x}\neq\vec{y}$.) From the point of view of electrostatics
this a finite collection of closed curves carrying either positive
or negative line charges. Due to the presence of the denominators
in the above formula for $\tau$ it is clear that there
is a negative real number $-t_0$ of very large absolute value such that the
equipotential surface $\tau=-t_0$ breaks up into $m$ connected surfaces
located each in the vicinity of one of the ``incoming curves'' $x^-_i(S^1)$.
The following intuitive argument
shows that each connected piece is diffeomorphic
to a Riemann surface of genus equal to the number of loops which are
generated by the self-intersections. In other words: the level surface piece
near such a curve looks like a two-sphere if the curve is just a point,
like a two-torus if the curve is immersed with no self-intersection,
like a surface of genus two if the curve looks like the figure 8, and so
forth.
Indeed, since $x^-_i(S^1)$ is compact we can
cover it by a finite number of balls of diameter $L$ where $L$ can be chosen
arbitrarily small. Now suppose that the map $x^-_i$ is well-behaved enough
so that we can approximate it by a sequence of straight pieces of wire
each carrying a constant line charge equal to the mean value of
$\lambda^-_i$ over the piece of the curve. Formula (\ref{wireeqn}) gives
an approximate solution of the potential $\tau$. From this formula
one can infer that if the distance of the point $\vec{x}$ to the curve
is very small compared to $L$ the potential is approximately proportional
to the natural logarithm of the distance. Hence for ``very early''
values of $\tau$ the connected components of the equipotential surface
are approximated by surfaces of constant distance from a given curve
which visibly have the required topology. One has the same
argumentation for very large or ``very late'' values of $\tau$: here
the connected components of the equipotential surfaces are certain Riemann
surfaces centered around the ``outgoing curves'' $x^+_j(S^1)$.

We can interpret this geometrical picture as a theory for propagating
membranes with prescribed change of topology: if the time equals $-\infty$
one has $m$ incoming membranes shrunk to one-dimensional objects
whose homotopy and initial ``shape'' are encoded in the curves $\vec{x}^-_i$
with line charge functions $\lambda^-_i$. These $m$ membranes
will grow bigger and possibly melt together (which can be interpreted as
interaction) and for large positive time values the surface of equal time
decomposes again into $n$ connected components which will eventually
center around the outgoing curves $\vec{x}^+_j$ with final ``shape'' described
by $\lambda^+_j$ and will be equal to these curves for $t=+\infty$.

In order to get more general solutions one has to solve the Dirichlet
problem for the potential $\tau$ outside $m$ conducting Riemann surfaces
held at value $-t_0$ and $n$ conducting Riemann surfaces held at value
$+t_0$ for a positive real number $t_0$ whose solution in principle exists
and is unique.

In order to get explicit solutions for the model (\ref{Nahm}) in
higher dimensionsional Euclidean space $\Real^{M+1}$ we can easily
generalize the above construction
(\ref{Nahmsolution}): we have to replace the curves
$\vec{x}^{\mp}_i(\vp)$ and by (at most) $M-1$ dimensional surfaces
$\vec{x}^{\mp}_i(\vp^1,\ldots,\vp^{M-1})$
equipped with higherdimensional ``line charge'' functions and the line
integral over
$1/|\vec{x}-\vec{x}^-_i(\vp)|$ by a surface integral (of codimension two)
over
$1/|\vec{x}-\vec{x}^-_i(\vp^1,\ldots,\vp^{M-1})|^{M-1}$.
More generally, one has to explicitly solve the higher dimensional analogue
of the above-mentioned Dirichlet problem
whose solution in principle exists and is unique.

Finally one should point out that (\ref{Nahm})
may also be written as
\be
   \dot{x}^i=\int_{\Sigma}\!d^M\tilde{\vp}~H^{ij}(\vp,\tilde{\vp})
                 \f{\delta W}{\delta x^i(\tilde{\vp})}[x]
\ee
with
\bea
     H^{ij}(\vp,\tilde{\vp}) & := &
         \f{\delta^{ij}\delta^{M}(\vp,\tilde{\vp})}{\rho(\vp)} \\
     W[x]                    & := &
     \f{1}{M+1}\int_{\Sigma}\!d^M\vp~x^i(\vp)\epsilon_{ii_1\cdots i_M}
                \epsilon^{r_1\cdots r_M}\f{1}{M!}
                \p_{r_1}x^{i_1}(\vp)\cdots\p_{r_M}x^{i_M}(\vp) \nonumber \\
                             &    &
\eea
in $\Real^ n$, i.e. are gradient flows with respect to the volume functional
$W$ and the metric $H^{ij}(\vp,\tilde{\vp})$.

\vspace{0.4cm}

{\bf 4.} Consider now a Hamiltonian field theory with
fields $(x,p)$ from $\Sigma$ into the cotangent bundle $T^*\Nf$ of $\Nf$
where the conjugate momentum field $p$ is supposed to be densitized.
\footnote{Although we shall be always working in canonical $x^i$ and $p_i$
  co-ordinates (a bundle chart) on $T^*\Nf$ this has a global meaning:
  the pair $(x,p)$ is a smooth vector bundle homomorphism
  $\wedge^M T\Sigma\ra T^*\Nf$ over
  the map $x:\Sigma\ra\Nf$, i.e.
  $x\circ\wedge^M\tau_\Sigma=\tau^*_\Nf \circ (x,p)$ with the obvious
  bundle projections.}
Let the Hamiltonian be of the form
\be \label{Ham}
   H(x,p):=\int_\Sigma\!d^M\vp~\sqrt{g}[x] h(\f{|p|}{\sqrt{g}[x]})
\ee
where $h$ is a smooth function of one variable,
$\sqrt{g}:=\sqrt{g}[x]:=\sqrt{\det g_{rs}}$ is defined as in (\ref{smallg}),
$p^2:=p_ip_j\zeta^{ij}(x),|p|:=\sqrt{p^2}$, and $u$ will be
short for $\f{|p|}{\sqrt{g}}$.
\footnote{In order to avoid clumsy notation
we shall henceforth suppress the arguments $\vp\in\Sigma$ in the integrals,
the field $x$ and its derivatives in $\sqrt{g}[x]$ and $n[x]$,
and the argument $u$ of $h$ and $h'$ from time to time.}

{}From the canonical Poisson structure
\be \lb{canPois}
  \{F,G\}=\int_\Sigma\!d^M\vp\big(
        \f{\delta F}{\delta x^i(\vp)}\f{\delta G}{\delta p_i(\vp)}
      - \f{\delta G}{\delta x^i(\vp)}\f{\delta F}{\delta p_i(\vp)} \big)
\ee
we obtain Hamiltonian equations of motion
\bea
  \f{\p x^i}{\p t}  & = & \zeta^{ij}(x)h'\f{p_j}{|p|}
                                                         \lb{dieqs} \\
  \f{\p p_i}{\p t}  & = & \zeta_{ij}(x)\p_r\big((h-uh')
                            \sqrt{g}g^{rs}\p_sx^j \big) \nonumber \\
          & & +\zeta_{ij}(x)(h-uh')\sqrt{g}g^{rs}\p_rx^k\p_sx^l
                        \Gamma^j_{kl}(x) \nonumber \\
          & & -\f{1}{2}\p_i\zeta^{jk}(x)h'\f{p_jp_k}{|p|}
\lb{dieps}
\eea
Defining
\be \lb{gendiff}
  C_r(x,p):=p_i\p_rx^i
\ee
(the infinitesimal generators of diffeomorphisms) it is easy to check
that (\ref{dieqs}) und (\ref{dieps}) imply
\be \lb{Ccons}
   \f{\p C_r}{\p t}=0
\ee
($H$ is invariant under diffeomorphisms $\phi:\Sigma\ra\Sigma$), as well
as
\be \lb{sechs}
  \f{\p}{\p t}\big( \sqrt{g}h \big)
                   =
             \p_r\big( (\f{hh'}{u}-h'^2)g^{rs}C_s(x,p)
                                          \big).
\ee
For solutions $(x(t),p(t))$ of the equations of motion satisfying
$C_r(x(0),p(0))=0$ the r.h.s. of eqn (\ref{sechs}) vanishes identically
(for all $t$ due to (\ref{Ccons})), hence (for such solutions)
\be \lb{dieDichte}
  \sqrt{g}[x(t)]h(\f{|p|(t)}{\sqrt{g}[x(t)]})=:\rho
\ee
with $\rho$ some {\em time independent} density on $\Sigma$.
Inverting (\ref{dieDichte}) to obtain $\f{|p|}{\sqrt{g}}$ as a
function of $\f{\sqrt{g}}{\rho}$, and noting that for $N-M=1$
the condition
\be \lb{derconstraint}
  C_r(x,p)=p_i\p_rx^i=0,~~~1\leq r\leq M
\ee
for $x$ an immersion means that $\zeta^{ij}p_j$ must lie in the direction
normal to the hypersurface defined by $\vp\mapsto x(t,\vp)$,
one sees that (\ref{dieqs}) can be rewritten in the form mentioned in
section 2 (see (\ref{surf}))
\be
   \f{\p x}{\p t}=\alpha(\f{\sqrt{g}}{\rho})n
\ee
with the functions $\alpha$ and $h$ being related via
\be
   \alpha(z)z^2\p_z(h^{-1}(\f{1}{z}))=-1.
\ee
In particular, for $h(z)=\sqrt{2z}$ we get $\alpha(z)=z$ (\ref{Nahm}) and
\be \lb{NahmHam}
    H[x,p]=\int_\Sigma\!d^M\vp~(4g[x]p^2)^{\f{1}{4}}.
\ee
For $h(z)=\sqrt{1+z^2}$ which corresponds to $\alpha(z)=\sqrt{1-z^2}$
we get the Hamiltonian
\be \lb{membHam}
   H[x,p]=\int_\Sigma\!d^M\vp~\sqrt{p^2+g[x]}
\ee
of the relativistic $M$-brane in $M+1$ dimensional Minkowski space.
$h(z)=z$ implying $\alpha(z)=1$ gives the Hamiltonian
\be
   H[x,p]=\int_\Sigma\!d^M\vp~\sqrt{p^2}
\ee
of the Optical model (\ref{Optical}).

Let us for a moment concentrate on (\ref{NahmHam}): Motivated by the
fact that the equation for the time function is linear (see \ref{harm})
and the observation \cite{Hop94}, made for $N=3$, that a Lax pair
formulation of $\p \vec{x}/\p t=\sqrt{g}\vec{n}$ implies the
time-independance of $\int\!Q(x(t,\vp^1,\vp^2)d\vp^1d\vp^2$
for harmonic polynomials $Q(x^1,x^2,x^3)$, one is led to the
conjecture that for all $N=M+1$, $\rho$, and $Q(x)$
any harmonic function of $x$,
\be
   {\bf Q}[x]:=\int_\Sigma\!d^M\vp~\rho ~Q(x)
\ee
will be independent of time if the hypersurface $\Sigma_t=x_t(\Sigma)$
evolves according to (\ref{Nahm}).
The proof is a trivial application of Gauss' Theorem: assuming that
for all $t$ the moving surface $\Sigma_t$ is a boundary of some
open set $V_t$ of $\Nf$ and writing $dS_t^i$ for the surface element we get
\beas
     \f{\p {\bf Q}[x(t)]}{\p t}  & = &
          \int_\Sigma\!d^M\vp\rho~\p_iQ(x(t))\f{\p x^i}{\p t} \\
                         & = &
          \int_\Sigma\!d^M\vp\sqrt{g}[x(t)]~n^i[x(t)]\p_iQ(x(t)) \\
                         & = &
          \int_{\Sigma_t}\!dS_t^i~\p_iQ(x(t)) \\
                         & = &
          \int_{V_t}\!d^Nx\sqrt{\det(\zeta_{kl})}~\p_j(\zeta^{ji}\p_iQ) \\
                         & = &
          \int_{V_t}\!d^Nx\sqrt{\det(\zeta_{kl})}~\Delta Q  ~=~      0.
\eeas
In the Hamiltonian formulation (\ref{NahmHam}) the nondynamical density
$\rho(\vp)$ is simply replaced by the Hamiltonian density
\be \lb{calH}
   {\cal H}[x,p](\vp) :=\sqrt{g}[x](\vp)~h(\f{|p|(\vp)}{\sqrt{g}[x](\vp)}),
\ee
i.e. ${\cal H}[x,p]=(g[x]p^2)^{\f{1}{4}}$ in the case of (\ref{NahmHam}).
Indeed, for any harmonic $Q$ the following functionals on phase space,
\be \lb{charge}
   {\bf Q}[x,p]:=\int_\Sigma\!d^M\vp~(gp^2)^{\f{1}{4}}Q(x),
\ee
will be time-independent for solutions $(x(t),p(t))$ satisfying
(\ref{dieqs}), (\ref{dieps}), and the constraint (\ref{derconstraint}).

Do the quantities (\ref{charge}) Poisson-commute (on the reduced space)?
The answer is
`yes' (in fact they form an infinite, presumably complete
set of Poisson-commuting, reparametrisation invariant
charges, thus providing the solution of a non-trivial diffeomorphism
invariant field theory)--but let us first make some observations valid
for all theories of the general form (\ref{Ham}), i.e. arbitrary
$h$, $N$, $M$, and $\zeta$: With the canonical Poisson structure
(\ref{canPois})
one has the following Poisson brackets where $f,\tilde{f}$ are arbitrary
smooth real-valued functions on $\Sigma$ and $X,Y$ are arbitrary vector
fields on $\Sigma$:
\bea
  \{\int_\Sigma\!d^M\vp ~X^rC_r,\int_\Sigma\!d^M\vp~ Y^sC_s\}
               & = &
                    \int_\Sigma\!d^M\vp~ [X,Y]^r C_r  \lb{CC}\\
 \{\int_\Sigma\!d^M\vp~ f{\cal H},
             \int_\Sigma\!d^M\vp~ \tilde{f}{\cal H}\}
               & = &
   \int_\Sigma\!d^M\vp ~\big(\f{hh'}{u}-h'^2\big)
                                           \nonumber \\
               &   & ~~~g^{rs}(f\p_s\tilde{f}
                      -\tilde{f}\p_s f){\cal H} \nonumber \\
               &   &      \lb{HH}   \\
  \{ \int_\Sigma\!d^M\vp~ f{\cal H},\int_\Sigma\!d^M\vp ~Y^s C_s \}
               & = &
          -\int_\Sigma\!d^M\vp~Y^s\p_s f {\cal H} \lb{CH}
\eea
where $[X,Y]^r$ denotes the Lie bracket $\p_sY^rX^s-\p_sX^rY^s$ of $X$
and $Y$.

While (\ref{CC}), together with (\ref{Ccons}), imply that in principle
the Hamiltonian theory (\ref{Ham}) may be reduced to a Hamiltonian
theory on a diffeomorphism-invariant phase space (involving only
$N-M$ fields and $N-M$ momenta) one should note that on the reduced
phase space the density $\cal H$ (\ref{calH}) can no longer be defined,
unless integrated over particular functions. This explains that although
$\dot{\cal H}=0$ (\ref{sechs}) for solutions of (\ref{dieqs}) and
(\ref{dieps}) satisfying the constraint (\ref{Ccons}), the Dirac bracket
of $\cal H$ with $H$ (see e.g. \cite{YITP} for the case (\ref{membHam}))
would not give zero, due to (\ref{CH}). In particular, $\cal H$ is
{\em not} a conserved quantity in the Hamiltonian theory.

The charges ${\bf Q}$ (\ref{charge}), however,
{\em are} conserved on the reduced phase space:
${\bf Q}[x,p]=\int_\Sigma\!d^M\vp~{\cal H}Q(x)$ commutes with $C_r$,
\bea \lb{QC}
   \{ \int_\Sigma\!d^M\vp~{\cal H}Q(x),\int_\Sigma\!d^M\vp~Y^rC_r\}
                & = & -\int_\Sigma\!d^M\vp~Y^r\p_rQ(x){\cal H}
                                               \nonumber \\
                &   & +\int_\Sigma\!d^M\vp~{\cal H}\p_iQ(x)Y^r\p_rx^i = 0
\eea
(irrespective of $Q$, $h$, and $N-M$; just use (\ref{CH}) with
$\tilde{f}=Q(x)$ for the first term on the r.h.s of (\ref{QC})), as well
as with $H$ (provided $C_r=0$, $N=M+1$, and $hh'=const$).
Moreover,
\bea
  \lefteqn{ \{ \int_\Sigma\!d^M\vp~{\cal H}Q(x),
                 \int_\Sigma\!d^M\vp~{\cal H}\tilde{Q}(x)  \} }
                                             \nonumber \\
                  & &
              =\int_\Sigma\!d^M\vp~\big( \f{hh'}{u}-h'^2 \big)
                    g^{rs}\big( Q(x)\p_s\tilde{Q}(x)-
                                \tilde{Q}(x)\p_sQ(x)\big)C_r \nonumber \\
                  & & ~~~~-\int_\Sigma\!d^M\vp \sqrt{g}[x]~h'h
                      \f{p_i}{|p|}\zeta^{ij}(x)\big( Q(x)\p_j\tilde{Q}(x)-
                                \tilde{Q}(x)\p_jQ(x)\big) \nonumber \\
                  & & ={\rm const.}\int_\Sigma\!d^M\vp \sqrt{g}[x]~
                          n^j[x]\big( Q(x)\p_j\tilde{Q}(x)-
                                \tilde{Q}(x)\p_jQ(x)\big) \nonumber \\
                  & & ={\rm const.}\int_{V_t}\!d^Nx \sqrt{\zeta_{kl}}~
                          \p_i\big( \zeta^{ij}(Q\p_j\tilde{Q}-
                                \tilde{Q}\p_jQ)\big) = 0 \label{QQtil}
\eea
where again we have assumed that $\Sigma_t$ is the boundary of $V_t$.
In (\ref{QQtil}), the first equality is general (using (\ref{HH}),
with $f=Q(x),\tilde{f}=\tilde{Q}(x)$, for the first term), the second
equality holds if $C_r=0$ ($1\leq r\leq M$) and $N=M+1$ (so that
$\zeta^{ij}p_j$ is proportional to the surface normal), and the last
equality holds if $Q$ and $\tilde{Q}$ are harmonic functions
(it actually holds as long as $\tilde{Q}\Delta Q-Q\Delta\tilde{Q}=0$,
which is a much weaker requirement).

The preceding observation has far-reaching consequences. It not only
proves the Poisson-commutativity of the conserved charges $Q$ given
in (\ref{charge}) (--hence the integrability of
(\ref{NahmHam}); the freedom of choosing $Q(x)$ in $\Real^3$ e.g.
as $\sum_{l=0,|m|\leq l}^{\infty}q_{lm}Y_{lm}(\vec{x})$ with
$Y_{lm}(\vec{x})$ being the solid spherical harmonics, neatly
matches the degrees of freedom to be expected from the one single
field that is left over after the Hamiltonian reduction--) but
--confirming that integrable theories come in hierarchies--
shows that {\em any} of the charges (\ref{charge}) with $\Delta Q=0$
may be considered as the Hamiltonian of another diffeomorphism
invariant field theory (possessing (\ref{charge}) as commuting
conserved charges).


Due to
\be
  \int_\Sigma\!d^M\vp~Q(x)(4g[x]p^2)^{\f{1}{4}}
              = \int_\Sigma\!d^M\vp~(4\tilde{g}[x]\tilde{p}^2)^{\f{1}{4}}
              =:\tilde{H}[x,p]
\ee
corresponding to $\tilde{Q}(x)=1$ for a Riemannian manifold
$(\Nf,\tilde{\zeta})$ with conformally equivalent metric
\be \lb{conformalchange}
   \tilde{\zeta}_{ij}(x):=(Q(x))^{\f{4}{N-2}}\zeta_{ij}(x)
\ee
(implying $\sqrt{\tilde{g}}=(Q(x))^{2\f{N-1}{N-2}}\sqrt{g}$,
$\tilde{p}^2=\tilde{p}_i\tilde{p}_j\tilde{\zeta}^{ij}
 =p_ip_j\zeta^{ij}Q^{-\f{4}{N-2}}$; $N\neq 2$; the very special case
$N=2$ can be dealt with separately), and the fact that under the
conformal change (\ref{conformalchange}) the scalar curvature
changes according to
\bea
   \tilde{R} & = & Q^{-\f{4}{N-2}}\big( R-4\f{N-1}{N-2}\f{\Delta Q}{Q}
                                 \big) \\
             & = & Q^{-\f{4}{N-2}} R
\eea
one sees that each hierarchy (of integrable Hamiltonian systems, resp.
harmonic time-functions) consists of hypersurface motions in Riemannian
manifolds having conformally equivalent metrics and satisfying
\be
    \tilde{R}\tilde{\zeta_{ij}}=R\zeta_{ij}.
\ee
This intriguing observation should have many important consequences.
Taking e.g. $\Nf =\Real^3$ and $\zeta_{ij}=\delta_{ij}$ the above suggests
that the solutions of the linear, but nontrivial, equation
\be
  \vec{\nabla}\cdot(Q^2(\vec{x})\vec{\nabla}\tau (\vec{x}))=0
\ee
resp. the corresponding nonlinear equations
\be
  \f{\p \vec{x}}{\p t}=Q^2(\vec{x})
                \f{\sqrt{\det(\p_r\vec{x}\cdot\p_s\vec{x})}}{\rho}
                                  \vec{n}
\ee
are related to
\be
  \vec{\nabla}^2\tau (\vec{x})=0
\ee
resp.
\be
  \f{\p \vec{x}}{\p t}=
                \f{\sqrt{\det(\p_r\vec{x}\cdot\p_s\vec{x})}}{\rho}
                                  \vec{n}
\ee
via some generalized B\"acklund transformation for all $Q(\vec{x})$
satisfying
\be
  \vec{\nabla}^2Q(\vec{x})=0.
\ee

\vspace{0.4cm}

\noindent {\bf\Large Acknowledgment}

\vspace{0.2cm}

\noindent The authors would like to thank J.~Fr\"ohlich, D.~Giulini,
K.~Happle, and M.~Struwe for
valuable discussions, C.~Kiefer for pointing out ref. \cite{BM88},
and the
Physics Department of the ETH Z\"urich for friendly hospitality.


\begin{thebibliography}{99}

\bibitem{stern} K.~A.~Brakke: The motion of a Surface by its mean curvature.
                Mathematical Notes, Princeton University Press, Princeton
                N.~J.~1978. \newline
                G.~Huisken: Flow by mean curvature of convex surfaces into
                spheres. J.~Diff.~Geom. {\bf 20} (1984), 237. \newline
                L.~Evans and J.~Spruck: Motion of level sets by mean
                curvature. J.~Diff.~Geom. {\bf 33} (1991), 635. \newline
                Y.~G.~Chen, Y.~Giga, S.~Goto: Uniqueness and existence
                of viscosity solutions of generalized mean curvature
                flow equations. J.~Diff.~Geom. {\bf 33} (1991), 749.

\bibitem{sternstern} P.~Pelc\'e: Dynamics of Curved Fronts.
                Academic Press, New York, 1988. \newline
                S.~Osher and J.~A.~Sethian: J.~Comput.~Physics {\bf 79}
                (1988), 12.

\bibitem{BM88} C.~Bona and J.~Mass\'o: Harmonic synchronisations
 of spacetime. Phys.~Rev. {\bf D 38} (1988), 2419-2422.


\bibitem{BH94} M.~Bordemann and J.~Hoppe: The Dynamics of Relativistic
 Membranes II: Nonlinear Waves and covariantly reduced membrane
 equations.
 Phys.~Lett.~{\bf B 325} (1994), 359-365.

\bibitem{War90} E.~G.~Floratos and G.~K.~Leontaris:
       Phys.~Lett.~{\bf B 223} (1989), 153. \newline
       R.~S.~Ward: Linearization of the $SU(\infty)$ Nahm
       Equations. Phys.~Lett. {\bf B 234} (1990), 81-84.

%

\bibitem{Hop94} J.~Hoppe: Surface motions and fluid dynamics.
 Phys.~Lett.~{B 335} (1994), 41-44.

\bibitem{YITP} J.~Hoppe: Canonical $3+1$ Description of Relativistic
     Membranes, Yukawa Institute Preprint YITP/K-1079.

\bibitem{what?} N.~Hitchin: On the construction of Monopoles.
     Comm.~Math.~Phys. {\bf 89} (1989), 145-190.

\bibitem{NHW95} K.~Nakayama, J.~Hoppe, M.~Wadati: On the Level-Set
 Formulation of Geometrical Models. J. of the Physical Society of Japan
 {\bf 64}, No.2 (1995), 403-407.

\bibitem{NW95} K.~Nakayama, M.~Wadati: Reaction-Diffusion System in a
 Curved Space and the KPZ Equation. J. of the Physical Society of Japan
 {\bf 64}, No.5 (1995), 1501-1505.

\bibitem{Mos65} J.~Moser: On the volume element on a manifold.
Trans.~Am.~Math.~Soc. {\bf 120} (1965), 286-294.

\bibitem{Jack75} J.~D.~Jackson: Classical Electrodynamics. $2^{\rm nd}$
 edition, New York, 1975.


\end{thebibliography}
\end{document}